\documentstyle[12pt,epsf]{article}
\global\let\epsfloaded=Y 

\catcode`@=12
\newcommand{\be}{\begin{equation}}
\newcommand{\ee}{\end{equation}}
\newcommand{\beq}{\begin{eqnarray}}
\newcommand{\eeq}{\end{eqnarray}}

\begin{document}
\begin{titlepage}

\rightline{\vbox{\halign{&#\hfil\cr
&NTUTH-97-11\cr
&August 1997\cr}}}
\vspace{0.2in}

\vfill

\begin{center}
{\Large \bf  Massiveness of Glueballs as Origin of the OZI Rule
}
\vfill
        {\bf Wei-Shu Hou and Cheng-Ying Ko}\\
        {Department of Physics, National Taiwan University}\\
        {Taipei, Taiwan 10764, R.O.C.}\\
\end{center}
\vfill
\begin{abstract}
The heaviness of the glueball mass scale is suggested as 
the source of the OZI rule at low energy. 
The $J/\psi \to \rho\pi$ decay ``anomaly" implies 
the vector glueball $O$ has mass $m_O \approx m_{J/\psi}$.
Such a heavy mass is supported by other glueball studies.
Glueball-meson matrix elements turn out to be not suppressed at all
at the 1 GeV scale, and a simple and intuitive picture emerges 
which is consistent with the Gell-Mann--Okubo mass formula 
as well as the measured sign of $\phi$-$\omega$ mixing.
The suppression of glueball mediated $\bar q_iq_i \longleftrightarrow \bar q_jq_j$ 
transitions and the cancellation mechanism in two-step meson rescatterings
are viewed as related by duality.
Extensions to the $2^{++}$, $3^{--}$ meson sectors,
and failure for $0^{\pm +}$ mesons are also discussed.
\end{abstract}
\vfill
\end{titlepage}
%
%

\vskip0.7cm
\noindent{\large\bf 1. Introduction}
\vskip0.2cm

When applying the Gell-Mann--Okubo mass formula to vector mesons,
it was found that the physical $m_\omega$ differed significantly from $m_{\omega_8}$.
This lead to the proposal of $\phi$-$\omega$ mixing \cite{Sakurai},
which turned out to be near ideal.
To explain why $\phi \rightarrow 3\pi $ is so suppressed as compared to
$\phi \rightarrow K \bar K$ and $\omega \to 3\pi $,
Okubo \cite{Okubo}, Zweig \cite{Zweig}, and Iizuka \cite{Iizuka}
independently suggested that strong processes in which the final
states can only be reached through $q$ and $\bar q$ annihilations 
(disconnected quark lines) are suppressed. 

The decay rate of $\phi \rightarrow 3\pi$ is not zero, hence the OZI rule is not exact, 
but since this decay rate is small, the violation of the OZI rule is soft. 
The purpose of this paper is 
to try to give a dynamical explanation of the OZI rule and its violation.
The dynamical sources of OZI violation are 
usually classified as from two origins: 
(1)  $q\bar q$ annihilations into multigluon intermediate states, and
(2) unitarity correction effects involving (physical or virtual) 
rescattering of $q\bar q$ hadronic states. 
At $\omega$, $\phi$ energies, 
the relative importance of these two sources is still an open question.

At J/$\psi $ energy scale, the dynamical origin of OZI violation
is understood as a manifestation \cite{AP} of asymptotic freedom
(class (1)),
which allows us to calculate OZI forbidden transition rates perturbatively. 
The question is how this picture can be extended \cite{DGG,FM} down to 
the $\omega$ and $\phi$ scale.
On the other hand, as championed by Lipkin \cite{Lipkin},
if one takes into account processes involving 
hadronic intermediate states (class (2)), 
e.g. $\phi \rightarrow K \bar K \rightarrow \rho \pi $ \cite{KKrhopi} or 
$\psi \rightarrow D \bar D \rightarrow \rho \pi $, 
OZI violation still needs to be understood. 
In these two-step transitions, each step is OZI allowed so 
the transition amplitude may not be small. 
It is suggested that cancellations \cite{SWSL} between different contributions
may result from conditions imposed by SU$(3)$ flavor symmetry,
nonet symmetry and exchange degeneracy \cite{Lipkin}. 
Following Lipkin's discussions, Geiger and Isgur \cite{GI} have
calculated these two-step transition amplitudes explicitly, using
closure and spectator approximations and the $^3P_0$ $q\bar q$ pair
creation model. The results confirm that the cancellation mechanism
indeed happens in detail.

The results after cancellation of two-step transitions could still
dominate the transition amplitude. 
But this would not be predictive because of its complexity
(some times up to $10^4$ terms \cite{GI}),
nor does it constitute a fundamental explanation of the OZI rule.
In this paper we assume that the cancellation is exact and 
consider gluonic intermediate states only,
in particular empahsizing the role of gluonic bound states, i.e. glueballs. 
This will lead to a simpler and more intuitive picture for 
the dynamical origin of the OZI rule and its violation. 
We shall use vector mesons as the prime example, 
then extend to $2^{++}$ and $3^{--}$ nonets.

\vskip0.7cm
\noindent{\large\bf 2. Vector Glueball Mass and Charmonium Decay}
\vskip0.2cm

The discovery of J/$\psi$ revived interests in the OZI rule in the 1970's.
In 1975, Freund and Nambu (FN) \cite{FN} suggested that
the breaking of the OZI rule in vector meson decays could be understood as due to
the mixing of $\omega$, $\phi$, and $J/\psi$ mesons with a new
SU$(4)$-singlet meson $O$, viewed as a ``Pomeron daughter".
Denoting the dimension two $O$-$V$ ($V=\omega$, $\phi$, $J/\psi$) transition amplitude 
as $f_{OV}$, one has
\be
   f_{O\psi}=f_{O\phi}= \frac{1}{\sqrt{2}}\, f_{O\omega}\equiv f.
\label{eq:f}
\ee
FN used dual dynamics to predict that
$m_O\sim 1.4$--$1.8$ GeV. Taking $f$ in Eq. (\ref{eq:f}) to be constant,
their approach failed to predict $J/\psi \to \rho\pi$ decay correctly.

By 1982, the so-called $J/\psi$ vs. $\psi^\prime$  decay anomaly appeared \cite{Franklin}.
Normally, one expects $J/\psi,\; \psi^\prime \rightarrow 3g\rightarrow X$ to
differ only in the charmonium wave function at the origin, hence the 
ratio of branching ratios is expected to follow the so-called $15\%$ rule,
\be
     \frac{{\cal B}(\psi^\prime \rightarrow X)}{{\cal B}(J/\psi \rightarrow X)} \simeq
    \frac{{\cal B}(\psi^\prime \rightarrow e^+e^-)}{{\cal B}(J/\psi \rightarrow e^+e^-)}
    \simeq 15\%
\label{eq:fifteen}
\ee
which appears to hold for general $X$. 
However, although $\rho\pi$ and $K^* \bar K$ decays are quite prominent 
($\sim 1\%$) for $J/\psi$, they were not seen in $\psi^\prime$ decay at all \cite{Franklin}.
To explain this anomaly, 
Hou and Soni (HS) \cite{HS} took $f = f(q^2)$ and 
proposed a resonance enchancement model,
viewing $O$ as the lowest lying vector glueball.
Assuming: i) $J/\psi \rightarrow O \rightarrow \rho\pi \gg
J/\psi\rightarrow ggg \rightarrow \rho\pi$,
ii) $J/\psi \rightarrow O\rightarrow$ other 
       $\ll J/\psi \rightarrow ggg \rightarrow$ other,
iii) $\psi^\prime\rightarrow O \rightarrow$ ANY 
       $\ll \psi^\prime \rightarrow ggg \rightarrow$ ANY, 
where $ggg$ stands for three-gluon continuum states,
one finds
\be
   \frac{\Gamma(\psi^\prime\rightarrow O \rightarrow \rho\pi)}
        {\Gamma(J/\psi\rightarrow O \rightarrow \rho\pi)}
   \simeq\left( \frac{m^2_\psi - m^2_O}{m^2_{\psi^\prime} - m^2_O}\right)^2
     \frac{f^2_{O\psi^\prime}}{f^2_{O\psi}},
\label{eq:polemodel}
\ee
where the energy denominator is the main enhancement factor for $J/\psi$.
Hence, as the anomaly deepened in 1986, the only way out was to have 
$m_O\simeq m_{J/\psi}$, as pointed out by 
Brodsky, Lepage and Tuan (BLT) \cite{BLT}. 
These authors also stressed that the $VP$ modes should otherwise
be suppressed by hadronic helicity conservation.

By 1996, the  ``$15\%$ rule" has been confirmed for \cite{Gu}
$p \bar p, p \bar p+n\pi, 5\pi, 7\pi, b_1\pi$ ($AP$) and $\phi f_0$ ($VS$) modes.
However, the $J/\psi$ anomaly persists for $VP$ modes,
\be
  {\cal B}(\psi^\prime \rightarrow \rho\pi) <2.9\times 10^{-5}, \ \
  {\cal B}(\psi^\prime \rightarrow K^{*+}K^-) <3.2\times 10^{-5},
\label{eq:VP}
\ee
while a similar situation starts to emerge for $VT$ modes such as 
$wf_2,\; \rho a_2$ and $K^* \bar K_2$. 
Concurrently, however, the degeneracy of $m_O \simeq m_{J/\psi}$
was challenged \cite{scan} by a BES energy scan of $J/\psi\to \rho\pi$,
which showed no sign of $O$ in the vicinity of $J/\psi$.
It turned out, however, that there was an analysis fault in Ref. \cite{scan}.
Upon closer scrutiny, it was found \cite{Hou} that 
$m_O\simeq m_{J/\psi}$ is not ruled out by the energy scan 
so long that $\Gamma_O \gg \Gamma_{J/\psi}$, which should be the case.
In fact, $O$ could hide even more easily in the radiative tail 
above the $J/\psi$ peak.
Taking $m_O \simeq 3200$ MeV for example, one gets 
rather plausibly \cite{Hou}
\be
4 \mbox{MeV} \ \raisebox{-.5ex}{\rlap{$\sim$}} \raisebox{.4ex}{$<$}\ \Gamma 
   \ \raisebox{-.5ex}{\rlap{$\sim$}} \raisebox{.4ex}{$<$}\ 30 \ \mbox{MeV, \ \ and \ \ }
\mbox{few}\, \% \ \raisebox{-.5ex}{\rlap{$\sim$}} \raisebox{.4ex}{$<$}\ 
   {\cal B}(O\to \rho\pi) \ \raisebox{-.5ex}{\rlap{$\sim$}} \raisebox{.4ex}{$<$}\ 25\, \%,
\label{eq:GammaO}
\ee
in contrast to the implicit need for $O\to \rho\pi$ dominance
in the original HS model \cite{HS}. One also obtains the mixing parameters
\be
f(m_{J/\psi}^2) \simeq 0.02 \mbox{\ GeV}^2, \ \ \ \sin\theta_{O\psi}\simeq 0.03.
\label{eq:f_psi}
\ee
The smallness of the $O$-$J/\psi$ transition strength $f(m_{J/\psi}^2)$ conforms well
with asymptotic freedom since it is $\propto \alpha_s^{3/2}$.
Because $f(m_{J/\psi}^2)$ is so small, one has a small mixing angle 
$\sin\theta_{O\psi}$ despite the proximity of $O$ and $J/\psi$ masses.
This implies that the $J/\psi$ mass shift due to mixing with $O$ is negligible
(at the sub-MeV level).

The above phenomenological arguments suggest an
$m_O$ value which is much larger than the prediction of FN, 
and turns out to be fortuitously \cite{Hou} close to $m_{J/\psi}$. 
Do we have other evidence to support this? 
Afterall, we have yet to establish any glueball state beyond doubt. 
We note, however, that recent experimental and lattice
studies are converging \cite{Landua} on $0^{++}$ and $2^{++}$ glueballs.
In the $0^{++}$ case, there is an excess of isoscalar mesons:
$f_0(1370)$, $f_0(1500)$ and $f_J(1710)$. 
Together with the $I=1/2$ and $1$ mesons $K^*_0(1430)$ and $a_0(1450)$, 
they cannot all fit into a $q\bar q$ nonet. 
Recent lattice calculations in the quenched approximation
predict the $0^{++}$ glueball mass to be $1600\pm 100$ MeV \cite{Landua},
which can fit either $f_0(1500)$ \cite{AC} or $f_J(1710)$ \cite{SVW}
as the $0^{++}$ scalar glueball $G$ while the other is dominantly $s\bar s$. 
It is likely that both states have large glueball admixtures. 
The situation seems cleaner though less established in the $2^{++}$ case. 
The  $even^{++}$ state $\xi(2230)$ is the glueball candidate \cite{xi},
which is very close in mass to the lattice expectation of $2400\pm 120$ MeV \cite{MP}. 
All these states are seen in $J/\psi\rightarrow\gamma+X$ transitions but
not seen in $\gamma\gamma$ production \cite{PDG}.

What does this have to do with the heaviness of $O$?
Note that $0^{++}$ and $2^{++}$ quantum numbers can be constructed 
from two gluons but the $1^{--}$ quantum number demands three gluons.
The $0^{++}$ quantum number is shared by the QCD vacuum 
hence is more complicated, but we could rather na\"\i vely 
scale from the $2^{++}$ glueball to the 3-gluon case,
which suggests $m_O$ to be in the ballpark of $m_{J/\psi}$.
Indeed, the original HS model \cite{HS} was motivated by 
$m_O$ expectations from a constituent gluon picture \cite{CS}.
The model predicts $m_{1^{--}}/m_{2^{++}} \simeq 1.5$ \cite{HS,CS,HL}, 
and assuming the lattice/experimental result on $\xi(2230)$,
it offers strong support \cite{Narison} for $m_O\sim m_{J/\psi}$.
Similar result of $m_{3^{--}}/m_{2^{++}} \simeq 1.5$ is obtained 
for the 3-gluon $3^{--}$ state \cite{HS,HL}.
Lattice studies of the $1^{--}$ glueball are unfortunately scarce and inconclusive, 
but it does turn out to be rather massive \cite{UKQCD}.
In the following, we shall take the glueball masses to be
$m_{0^{++}},\; m_{2^{++}} = 1600,\; 2230$ MeV and
$m_{1^{--}} \simeq m_{3^{--}} = 3200$ MeV respectively.
Note that, compared to 15 years ago, theses masses
are 30--40\% heavier.

\vskip0.7cm
\noindent{\large\bf 3. Vector Glueball and OZI Dynamics}
\vskip0.2cm

To understand the OZI rule in the glueball mediated picture,
we need to have better understanding of the strength of 
the transition amplitude $f$ at scales much lower than the
$m_{J/\psi}^2$ scale of Eq. (\ref{eq:f_psi}).
In the ideal mixing basis (limit of exact OZI rule), usually defined as
\be
   \phi_0 \equiv s\bar s, \ \ \
   \omega_0 \equiv \frac{1}{\sqrt{2}}(u\bar u+d\bar d) \equiv n\bar n,
\label{eq:ideal}
\ee
the transition amplitude $f$ of Eq. (\ref{eq:f}) induces the mixing
\beq
\phi &=& \phantom{-}\cos \delta\; \phi_0 +\sin \delta\; \omega_0  \nonumber \\
\omega &=& -\sin \delta\; \phi_0 +\cos \delta\; \omega_0,
\label{eq:phiomega}
\eeq
where the angle $\delta$ is the (OZI violating) deviation from ideal mixing. 
To see how $\delta$ relates to $f$, note that Eq. (\ref{eq:phiomega})
can be derived from the mass-squared matrix
\be
M^2=\left( \begin{array}{cc}
                 m^2_{\phi_0} &  T(\phi_0 \rightarrow \omega_0 ) \\
                 T(\omega_0 \rightarrow \phi_0 )   &   m^2_{\omega_0} \\
       \end{array}
    \right),
\ee
where the relation between $\delta$ and the off-diagonal transition amplitude
$T(\phi_0 \rightarrow \omega_0 )$ is
\be
\tan \delta =\frac{T(\phi_0 \rightarrow \omega_0)}
                {m^2_{\phi_0}-m^2_{\omega_0}}.
\label{eq:tandelta}   
\ee
The transition amplitude $T(\phi_0 \rightarrow \omega_0 )$
is expanded by a complete set of gluonic states $\vert x\rangle$,
\be
  T(\phi_0 \rightarrow \omega_0)=\sum_x\frac{\langle\omega_0\vert
     H_{int} \vert x\rangle\langle x\vert H_{int}\vert\phi_0\rangle}
     {m^2_{\phi_0}-m^2_x}.
\label{eq:completeset}
\ee
The matrix elements in the numerator, related to
the $f_{OV}$ defined earlier, are in principle calculable. 
The $O$ dominance picture amounts to saturating the sum by $O$, 
namely $T(\phi_0 \rightarrow \omega_0) \simeq 
\sqrt{2}f^2(m^2_{\phi_0})/(m^2_{\phi_0}-m^2_O)$
where
\be
f(m^2_{\phi_0})=\langle O\vert H_{int} \vert q\bar q\rangle
  \left.\phantom{1^t\over 2}\hskip -0.35cm \right\vert_{q^2 = m_{\phi_0}^2}.
\label{eq:fME}
\ee
$O$ dominance can be partially understood by the fact that 
$O$, $\omega$ and $\phi$ are all lowest lying states, 
hence the matrix elements involving excited glue states
are suppressed by both the numerator and the denominator.
We finally get
\be
 \tan \delta \simeq \frac{\sqrt{2}f^2(m^2_{\phi})}{(m^2_{\phi}-m^2_{\omega})
                (m^2_{\phi}-m^2_O)},
\label{eq:tandelta2}
\ee
where we have approximated $m_{\phi_0}$, $m_{\omega_0}$ 
by $m_\phi$, $m_\omega$.

From Eq. (\ref{eq:tandelta2}) and analogous relations for other mesons,
we can determine $f(m^2_{\rm had.})$ from meson and glueball masses 
and $\delta_{\rm expt.}$.
The latter is calculated by using the quadratic Gell-Mann--Okubo mass formula,
i.e. $\delta_{\rm expt.} = \delta_{\rm GMO}$, which gives results in good agreement 
with those extracted from decay data \cite{PDG,tornqvist}.
The results are given in Table 1.
The distinction between $\delta_{\rm GMO} = \delta_{n\bar n}$ 
and $\delta_{s\bar s}$ is explained in the next section.
We have not presented the results for $0^{-+}$ mesons since 
$\eta$-$\eta^\prime$ mixing is related to the famous U$_A(1)$ anomaly,
and does not follow the present formalism.
We took $s\bar s$ to be the $f_J(1710)$ state for the $0^{++}$ entry
as illustration, but the case is highly uncertain, 
and we do not pursue it further here.
\begin{table}[htb]
\vspace{5mm}
 \begin{center}
  \begin{tabular}{rcccrrc}\hline\hline
 $J^{PC}$ & $n\bar n$ & $s\bar s$ & Glueball & $\delta_{n\bar n}$ 
                              & $\delta_{s\bar s}$ & $f(m^2_{\rm had.})$ (GeV$^2$)\\ \hline
 $0^{-+}$ &  --- & --- &  --- & $-45^\circ$ & --- & --- \\
 $0^{++}$ & $f_0(1370)$ & $f_J(1710)$ & $G(1600)$ & $33^\circ$ & ---
                                                                       & $\sim$ 0.4 \\  \hline
 $1^{--}$ & $\omega(782)$ & $\phi(1020)$ & $O(3200)$ & $4^\circ$ & $-4^\circ$
                                                                           & $\simeq$ 0.4 \\
 $2^{++}$ & $f_2(1270)$ & $f_2^\prime(1525)$ & $\xi(2200)$ & $-5^\circ$ & $7^\circ$
                                                                                     & $\simeq$ 0.4 \\
 $3^{--}$ & $\omega_3(1670)$ & $\phi_3(1850)$ & $O_3(3200)$ & $-3^\circ$ & $3^\circ$
                                                                                        & $\simeq$ 0.4 \\ \hline\hline
  \end{tabular}
 \end{center}
\caption{
Values of mixing angle $\delta$ from Gell-Mann--Okubo mass formula
and glueball-quarkonium mixing strength $f$ from Eq. (13).
For $0^{\mp+}$ and the distinction of $\delta_{n\bar n}$ and $\delta_{s\bar s}$ 
see discussion in text.
}
\end{table}

The $\delta$ values for $1^{--}$, $2^{++}$, $3^{--}$ mesons are 
clearly different from $0^{\mp+}$, all exhibiting near-ideal mixing.
Quite remarkably, they all (including $0^{++}$) give
\be
f(m^2_{\rm had.}) \sim ( \mbox{0.6 GeV})^2,
\label{eq:f_had}  
\ee
{\it consistently}, where  $m_{\rm had.} \sim 0.8$--$1.7$ GeV.
Eq. (\ref{eq:f_had}) should not come as a surprise since,
as seen  from Eq. (\ref{eq:fME}), $f(m^2_{\rm had.})$ is 
a hadronic matrix element evaluated at normal hadronic scales.
Thus, in strong contrast to 
the smallness of $f(m^2_{J/\psi})$ in Eq. (\ref{eq:f_psi}) at $m^2_{J/\psi}$ scale,
$f(m^2_{\rm had.})$ is {\it not} suppressed for the $1^{--}$ \cite{Hou}, 
$2^{++}$, $3^{--}$ nonets. From this we infer that 
{\it the smallness of OZI violation for these nonets 
is not due to the smallness of the glueball mixing strength, 
but because of the heaviness of the lowest glueball state 
that mediates the $n\bar n$-$s\bar s$ mixing}.
Thus, it is the second factor in the denominator of
Eq. (\ref{eq:tandelta2}) rather than the transition matrix elements 
in the numerator that controls the violation of the OZI rule.
Even for the 2-gluon $2^{++}$ glueball, its mass scale is 
considerably above the  $f_2$-$f_2^\prime$ mixing scale.
In contrast, because of the proximity of glueball and
meson mass scalaes, as seen from Table 1,
the OZI rule is badly broken in the $0^{++}$ sector.
We illustrate our scenario pictorially in Fig. 1.

\vskip0.7cm
\noindent{\large\bf 4. Sign of $\phi$-$\omega$ Mixing}
\vskip0.2cm

Although the glueball mediation scenario could intuitively and
simply explain the OZI rule and its violation,
there is one potential difficulty that has to be faced.
Note that so far we have been cavalier with 
the sign of $\delta$ in Eq. (\ref{eq:tandelta2}).
Twenty years ago, Arafune, Fukugita and Oyanagi (AFO) \cite{Arafune} 
pointed out the importance of this sign,
which seems to imply that the dominant contribution to OZI rule violation 
comes from SU$_{\rm F}(3)$ octet intermediate states (rescattering!) 
rather than singlet states such as $O$. 
This threatens the foundation of the glueball mediation picture.
Let us investigate this problem.

Defining the octet and singlet states in the usual way as \cite{PDG}
\be
\omega_8 ={1\over \sqrt{6}}\, (u \bar u + d \bar d -2s \bar s),\ \ \ 
\omega_1 = {1\over \sqrt{3}}\, (u \bar u + d \bar d + s \bar s),
\label{eq:omega81}  
\ee
then, in the convention of Eq. (\ref{eq:ideal}), we have
\be
\phi_0 =  \sqrt{1\over 3}\, \omega_1 - \sqrt{2\over 3}\, \omega_8, \ \ \
  \omega_0  = \sqrt{2\over 3}\, \omega_1 + \sqrt{1\over 3}\, \omega_8.
\label{eq:idealdecomp}  
\ee
Since $O$ couples only to the singlet component of $\phi_0$ and $\omega_0$,
we have $\langle\omega_0|H_{int}|O\rangle\langle O|H_{int}|\phi_0\rangle > 0$.
Together with $m_O > m_{\phi_0}$, the glueball dominance model 
predicts that $\tan \delta < 0$ \cite{Arafune}.
This contradicts the result from the Gell-Mann--Okubo mass formula
listed under $\delta_{n\bar n}$ in Table 1,
as well as \cite{Arafune} the direct experimental probe 
of the sign of the mixing angle $\delta$ via 
the interference between $\phi$ and $\omega$ 
in the $e^+e^- \rightarrow \pi^+ \pi^- \pi^0$ \cite{Parrour} process.
The observation of constructive interference in the  energy domain of  
$m_\omega < E_{cm} < m_\phi$ implies $\tan \delta > 0$.

A seemingly trivial way out is to change the convention from Eq. (\ref{eq:ideal}) to
\be
   \phi_0 \equiv - s\bar s, \ \ \
   \omega_0 \equiv \frac{1}{\sqrt{2}}(u\bar u+d\bar d),
\label{eq:ideal-}
\ee
hence 
$\langle\omega_0|H_{int}|O\rangle\langle O|H_{int}|\phi_0\rangle < 0$
which na\"\i vely leads to $\tan \delta > 0$.
Things are not so simple, however, since instead of Eq. (\ref{eq:phiomega}),
one now has
\beq
\phi &=& \cos \delta\; \phi_0 -\sin \delta\; \omega_0  \nonumber \\
\omega &=& \sin \delta\; \phi_0 +\cos \delta\; \omega_0.
\label{eq:phiomega-}
\eeq
Following the same steps as Eqs. (\ref{eq:ideal}--\ref{eq:completeset}),
one again runs into a sign problem.

Since physics should be convention independent,
we should view the implications of the $e^+e^- \rightarrow \pi^+ \pi^- \pi^0$
experiment from this light.
Treating $\sin\delta$ as a perturbation,
the $\omega$ channel proceeds via $\gamma^*\to \omega_0$
and $\omega_0 \to  \pi^+ \pi^- \pi^0$,
whereas the $\phi$ channel proceeds via $\gamma^*\to \phi_0$ 
and $\omega_0 \to  \pi^+ \pi^- \pi^0$.
Since the $\gamma$-$\omega_0$ and $\gamma$-$\phi_0$ couplings
have opposite sign, 
and since the $\omega$, $\phi$ propagators are also of opposite sign
for $m_\omega < E_{cm} < m_\phi$, 
constructive interference implies that,
{\it regardless of conventions} for $\phi_0$ and $\omega_0$,
{\it the relative sign of  the projection of 
$\phi$ onto $\phi_0$ and $\omega_0$ is positive}.
This illustrates the problem one has with Eq. (\ref{eq:phiomega-}).

The solution to our problem lies in some interesting subtleties which are
usually overlooked in casual applications of the GMO formula,
namely, scale dependence and the value of $m^2_{11}$.
To be consistent, we now adopt the conventions of
the Particle Data Group (PDG).
In terms of the octet and singlet states of Eq. (\ref{eq:omega81}),
the physical states are defined as \cite{PDG}
\beq
  \omega &=& \phantom{-} \cos\theta_V\, \omega_1 + \sin\theta_V\, \omega_8, \nonumber \\
  \phi &=&   - \sin\theta_V\, \omega_1 + \cos\theta_V\, \omega_8.
\label{eq:decomp}  
\eeq
Ideal mixing is defined as 
$\theta_{\rm ideal} = \sin^{-1} (1/\sqrt{3}) \cong 35.3^\circ$,
that is
\be
\phi_0 =  -\sqrt{1\over 3}\, \omega_1 + \sqrt{2\over 3}\, \omega_8, \ \ \
  \omega_0  = \sqrt{2\over 3}\, \omega_1 + \sqrt{1\over 3}\, \omega_8.
\label{eq:idealdecomp-}  
\ee
Thus, the PDG convention is in fact that of Eq. (\ref{eq:ideal-}) 
rather than Eq. (\ref{eq:ideal}), i.e. $\phi_0 \equiv -s\bar s$.

Eq. (\ref{eq:decomp}) supposedly diagonalizes the 
singlet-octet mass-squared matrix, or
\beq
 \left( \begin{array}{rr}
            m^2_{11} & m^2_{18}\\
            m^2_{81} & m^2_{88}\\
           \end{array}
 \right)
 &=& \left( \begin{array}{rr}
             \cos\theta & -\sin\theta \\
             \sin\theta &  \cos\theta \\
           \end{array}
     \right)
     \left( \begin{array}{cc}
               m^2_{\omega} &     0        \\
                  0       & m^2_{\phi} \\
               \end{array}
     \right)
     \left( \begin{array}{rr}
             \cos\theta & \sin\theta \\
            -\sin\theta & \cos\theta \\
           \end{array}
     \right)  \nonumber   \\
 &=& \left( \begin{array}{ll}
     \cos^2\theta\, m^2_{\omega}+\sin^2\theta\, m_{\phi}^2 \ \, &
     \sin\theta \cos\theta\, (m_{\omega}^2 -m_{\phi}^2)      \\
     \sin\theta \cos\theta\, (m_{\omega}^2 -m_{\phi}^2)   &
     \sin^2\theta\, m^2_{\omega}+\cos^2\theta\, m_{\phi}^2  \\
            \end{array}
     \right).
\label{eq:diag}
\eeq
As is clear from Eq. (\ref{eq:decomp}), 
the physical $\omega$ state descends from $\omega_1$ 
after mixing in $\omega_8$ at the $\sin\theta_\omega$ level.
Thus, the $88$ element of Eq. (\ref{eq:diag}) gives the standard formula \cite{PDG}
\be
 \tan^2\theta_\omega = {m_{\phi}^2-m_{88}^2\over m_{88}^2-m_{\omega}^2}.
\label{eq:thetaomega}
\ee
From the GMO formula 
$m_{88}^2 =(4m_{K^*}^2 -m_{\rho}^2)/3 \simeq 0.87$ GeV$^2$,
we find $\theta_\omega \simeq 39^\circ$,
leading to $\delta_\omega \simeq + 4^\circ$ listed in Table 1,
hence $\tan\delta >0$.
But if one takes the convenient, usual {\it assumption} of a 
single rotation angle,
one gets $\phi = \cos \delta\; \phi_0 -\sin \delta\; \omega_0$
from Eq. (\ref{eq:phiomega-}), which contradicts 
the $\omega$-$\phi$ interference experiment as mentioned earlier.
We stress, however, that from the general point of view of QCD,
Eq. (\ref{eq:diag}) should be  evaluated at a given scale. 
Since there are {\it two} different physical scales to the problem, 
namely $m_\omega$ and $m_\phi$,
there is no reason why there should be just one mixing angle \cite{GI2}.
The two physical states $\omega$ and $\phi$ can have different
``$\phi$-$\omega$ mixing angles" without violating orthogonality. 

From the quark content of Eq. (\ref{eq:omega81}), 
the quark model gives (in the SU$(2)$ limit)
\be
m_{88}^2 = m_0 + {2\over 3}(m_u + 2 m_s), \ \ \ 
m_{11}^2 = m_0^\prime + {2\over 3}(2 m_u +  m_s).
\ee
Since they characterize different SU$(3)$ multiplets,
$m_0$ and $m_0^\prime$ are usually treated as in principle different.
However, since U$(3)$ or nonet symmetry is more apparent for vector mesons,
as reflected in $m_\omega \cong m_\rho$,
we take $m_0^\prime = m_0$.
This conforms with flavor independence of QCD and
is more reasonable than assuming the sum rule or trace relation
$m_\phi^2 + m_\omega^2 = m_{88}^2 + m_{11}^2$,
i.e. equal mixing angles for $\phi$ and $\omega$.
We therefore get a GMO-like formula 
\be
m^2_{11}= {1\over 3} (2m_{K^*}^2 +m_{\rho}^2) \simeq 0.73 \mbox{ GeV}^2,
\ee
which is smaller than the $m_{11}^2 \simeq 0.78$ GeV$^2$ value
from assuming equal mixing angles.
Eq.  (\ref{eq:diag}) now gives a second rotation angle
\be
 \tan^2\theta_\phi = {m_{11}^2-m_{\omega}^2\over m_{\phi}^2-m_{11}^2},
\label{eq:thetaphi}
\ee
leading to $\theta_\phi\simeq 31^\circ$, hence $\delta_\phi \simeq -4^\circ$, 
which is listed in Table 1 under $\delta_{s\bar s}$.
Thus, we see from Eq. (\ref{eq:phiomega-}) that the relative sign of  
the projection of $\phi$ onto $\phi_0$ and $\omega_0$ is now positive,
which conforms with the interference experiment.
Repeating the steps of Eqs. (\ref{eq:ideal}--\ref{eq:completeset}), 
one finds $\tan \delta_\phi =-T(\phi_0 \rightarrow \omega_0)/
(m^2_{\phi_0}-m^2_{\omega_0})$, which is 
consistent with $\delta_\phi < 0$ for $\phi_0 = -s\bar s$ \cite{peculiar}.
We have therefore constructed a completely consistent picture
for OZI violation.

The physical states $\omega$ and $\phi$ as defined by
Eqs. (\ref{eq:thetaomega}) and (\ref{eq:thetaphi}) are plotted in Fig. 2.
Although they appear nonorthogonal,
they are in fact orthogonal when evolved to the same scale \cite{peculiar}.
Applying the same formalism to $2^{++}$ and $3^{--}$ mesons,
we obtain the respective entries in Table 1, 
and the statevectors are also plotted in Fig. 2.
Note that the $\delta$ angles for $2^{++}$ and $3^{--}$ are opposite
those of the $1^{--}$ case.
The result for $2^{++}$ is also supported by experimental data.
For example, in the $\pi + N \to K^+ K^- +N$ reaction which probes 
the $\pi^+\pi^- \to f_2,\; f_2^\prime,\; a_2^0\to K^+ K^-$ reactions,
one observes destructive interference near the $f_2^\prime$ mass
over a broad $f_2$ and $a_2^0$ background \cite{Pawlicki}.
Since the effect is seen at the $m_{f_2^\prime}^2$ scale, 
one should use $\delta_{f_2^\prime} \simeq +7^\circ$ for
both $f_2^\prime$ and ``$f_2$" in an equation analogous to 
Eq. (\ref{eq:phiomega-}). 
Destructive intereference follows naturally.

We conclude that the sign of $\phi$-$\omega$ mixing,
as pointed out by AFO \cite{Arafune}, is indeed an important physical parameter.
But contrary to the assertion of AFO,
we find that our proposed glueball mediation scenario
for OZI violation in $1^{--}$, $2^{++}$ and $3^{--}$ meson mixings
is in harmony with both the GMO mass formulas and
direct experimental probes.

\vskip0.7cm
\noindent{\large\bf 5. Discussion and Conclusion}
\vskip0.2cm

This work started with the observation in Ref. \cite{Hou} that,
while $f(m_\phi^2) \sim 0.5$ GeV$^2$ as inferred from 
$\phi\to \rho\pi$ decay is not suppressed at all,
$f(m_\phi^2)/(m_O^2 - m_\phi^2)$ is very close to $\sin\delta_{\omega\phi}$.
This is now understood in terms of Eq. (\ref{eq:tandelta2}) since
it turns out that $f(m_\phi^2) \sim m_\phi^2 - m_\omega^2$.
We have extended the observation to $2^{++}$ and $3^{--}$ mesons. 
The empirical result of $f(m_{\rm had.}^2) \sim (600\ \mbox{\rm MeV})^2$
is rather reasonable, since it is nothing but  a strong interaction matrix element
of  mass-squared dimensions measured at normal hadronic scales
(Eq. (\ref{eq:fME})).

We note from Fig. 2 that the angle between $\phi$ and $\omega$
is superficially less than $90^\circ$ while for $2^{++}$ and $3^{--}$
the angles are superficially larger than $90^\circ$.
It is not clear what is the actual cause of this, but it is probably 
correlated with the fact that $m_\omega > m_\rho$ while
$m_{f_2} < m_{a_2}$ and $m_{\omega_3} < m_{\rho_3}$.
The latter case of $m_{n\bar n}(I=0) < m_{n\bar n}(I=1)$
seems more reasonable from the point of view of level repulsion
induced by quantum mixing.
Perhaps $m_\omega > m_\rho$ is due to mesonic rescattering effects
(class 2), but the results obtained by Geiger and Isgur with their 
central parameter values give also the wrong sign \cite{GI}.
Related to this, note also that 
$m_\phi^2 + m_\omega^2 > m_{\omega_8}^2 + m_{\omega_1}^2$,
while the situation is opposite for $2^{++}$ and $3^{--}$.
This should be compared to the $0^{-+}$ case where
$m_\eta^2 + m_{\eta^\prime}^2$ is considerably larger than
$m_{\eta_8}^2 + m_{\eta_1}^2$, usually considered as
a sign of the large glue content of $\eta$-$\eta^\prime$ mesons
caused by the U$_A(1)$ anomaly.

Since $1^{--}$ and $3^{--}$ are composed of three gluons while
$2^{++}$ is made of two, one may ask why $f(m_{f_2^\prime}^2)$ 
is not much larger than the other cases.
We observe that the $2^{++}$ $q\bar q$ mesons are $P$-wave, 
while the glueball is $S$-wave.
Thus, besides the heaviness of the corresponding glueball as seen in Fig. 1, 
configurational mismatch may be part of the cause for
near ideal mixing in the $2^{++}$ sector.
Then why is the $3^{--}$ mesons not 
even closer to ideal mixing since the $q\bar q$ mesons are $D$-wave?
The answer is in part that they are indeed so. 
We  further remark that the $3^{--}$  $S$-wave glueball 
can be viewed as composed of a gluon pair with total spin $2$ 
coupled to an additional spin 1 gluon.
In terms of spin content, this matches onto
the $q\bar q$ meson's $L=2$ and $S=1$.

One implication of near-ideal mixing of $1^{--}$, $2^{++}$ and $3^{--}$
mesons is that the corresponding glueballs $O$, $\xi$ and $O_3$
are relatively clean, that is, they have only small admixtures of $q\bar q$
(including $c\bar c$ \cite{Hou}) in them.
Their decays therefore proceed differently than $q\bar q$ mesons,
as reflected in the $\xi(2230)$.
The indirectly inferred $O$ width of Eq. (\ref{eq:GammaO})
is less than the already narrow $\xi$ width.
This narrowness reflects the fact that the lowest lying glueball
cannot decay via glueball channels,
and their decay to $q\bar q$ final states are OZI (or $\sqrt{\rm OZI}$?)
suppressed.
The ``cleaness" of these glueballs would make their identification
much easier, {\it once they are seen}.
In comparison, the $0^{++}$ glueball $G$ mixes strongly with 
the neighboring $n\bar n$ and $s\bar s$ states,
and will take much effort to establish its identity.

Finally, we think that the OZI suppression due to 
high glueball mass scale is at the root of the cancellation mechanism
studied by Lipkin and by Geiger and Isgur.
In short, the two results are related by duality.
The annihilation via all possible gluonic states (class 1) and
the rescattering via all possible (quark) hadronic states (class 2)
are dual to each other, much like the equivalence between 
parton level inclusive cross sections
and a complete set of allowed hardronic states.
If in QCD OZI suppression comes about 
because of the heaviness of the glueball mass scale
(single channel dominance is not necessary)
as compared to the $q\bar q$ meson mass scale,
such OZI suppression should then be automatically and strenuously
maintained in terms of hadronic intermediate states.
It is therefore both remarkable and understandable, then,
that the cancellation mechanism fails to be operative precisely
when the glueball mass scale is lowest,
namely the $0^{++}$ sector \cite{pbarp}.

In summary, we argue that the OZI rule for normal hadrons
such as $\phi$ and $\omega$ is due to the heaviness of the mediating
glueball mass scale rather than suppressed transition matrix elements,
unlike the case for $J/\psi$ decay.
Glueballs turn out to be rather heavy in QCD, 
but otherwise the effect is quantum mechanical.
This seems to resolve a long standing riddle: 
why is the OZI rule operative at the 1 GeV scale?
We find that it is necessary to treat the two physical nonet isoscalar mesons
as having different mixing angles.
If we treat the $m_q$-independent contributions to meson masses as
U$(3)$ invariant rather than SU$(3)$ invariant,
a simple, intuitive and consistent picture emerges
for glueball dominance of OZI violation.

\vskip0.7cm
\noindent{\large\bf Acknowledgements.}
WSH wishes to thank E. M. Henley, N. Isgur, L. F. Li, H. J. Lipkin, K. F. Liu 
and T. M. Yan for discussions.
This work is supported in part by NSC 86-2112-M-002-019
of the National Science Council of the R.O.C.

\newpage

{}

\begin{figure}
\let\picnaturalsize=N
\def\picsize{6in}
\def\picfilename{msquared.eps}
\ifx\nopictures Y\else{\ifx\epsfloaded Y\else\input epsf \fi
\global\let\epsfloaded=Y
%
%
\centerline{\ifx\picnaturalsize N\epsfxsize \picsize\fi \epsfbox{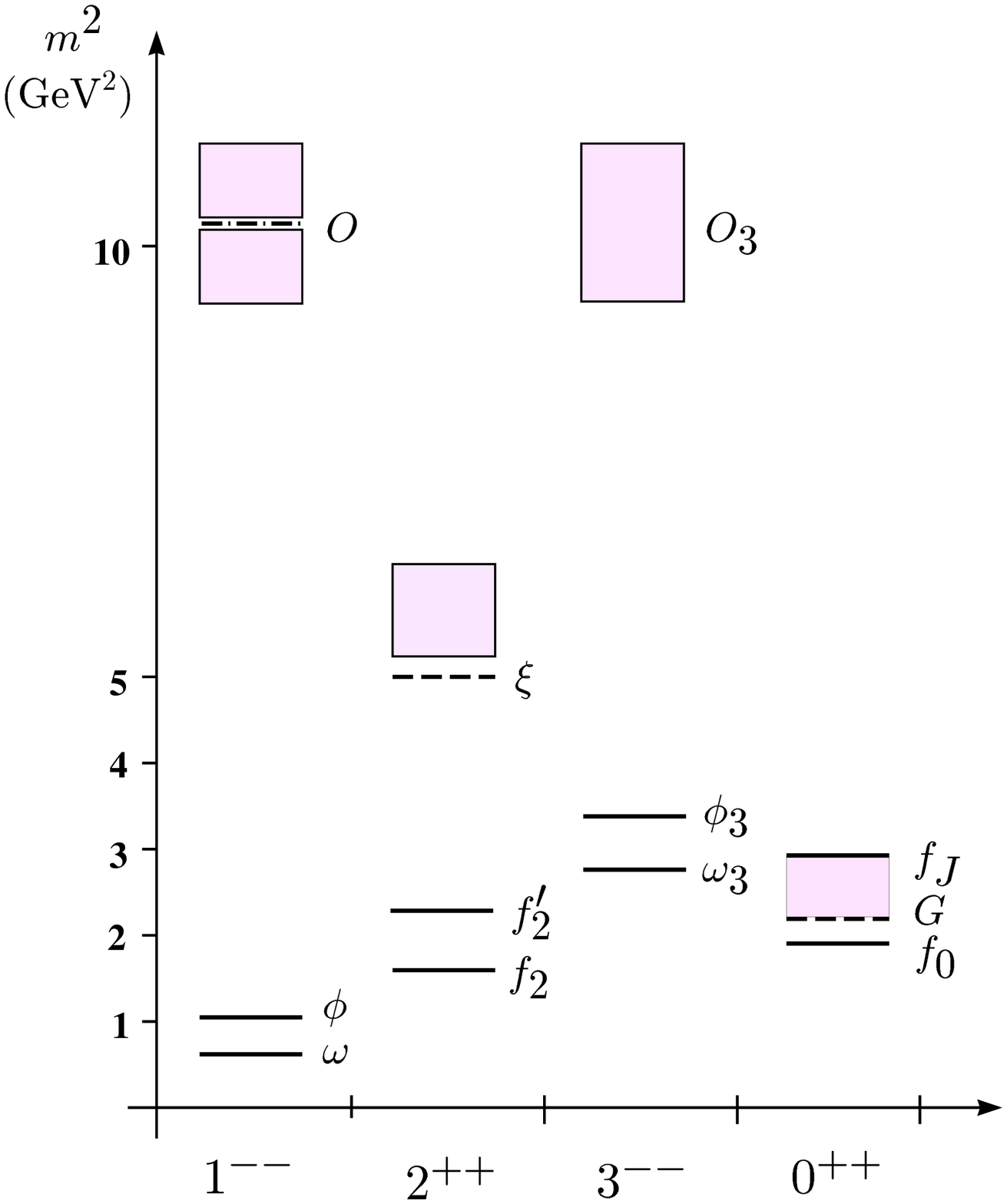}}}\fi
%
\vskip -0.2cm
  \caption{Mass-squared spectrum. 
   Solid lines denote mesons listed in the PDG 1996 booklet,
  dashed lines are glueball candidate states that needs further experimental
  confirmation,
  and dotdash line for $O$ stands for the $m_O$ value used in text.
   The shaded boxes are the expected glueball mass ranges from lattice or
   other estimates. 
   The glueball states are considerably heavier than the isoscalar mesons,
   except for the $0^{++}$ case.}
  \label{f1}
\end{figure}
\begin{figure}
\let\picnaturalsize=N
\def\picsize{6in}
\def\picfilename{phiomega.eps}
\ifx\nopictures Y\else{\ifx\epsfloaded Y\else\input epsf \fi
\global\let\epsfloaded=Y
%
%
\centerline{\ifx\picnaturalsize N\epsfxsize \picsize\fi \epsfbox{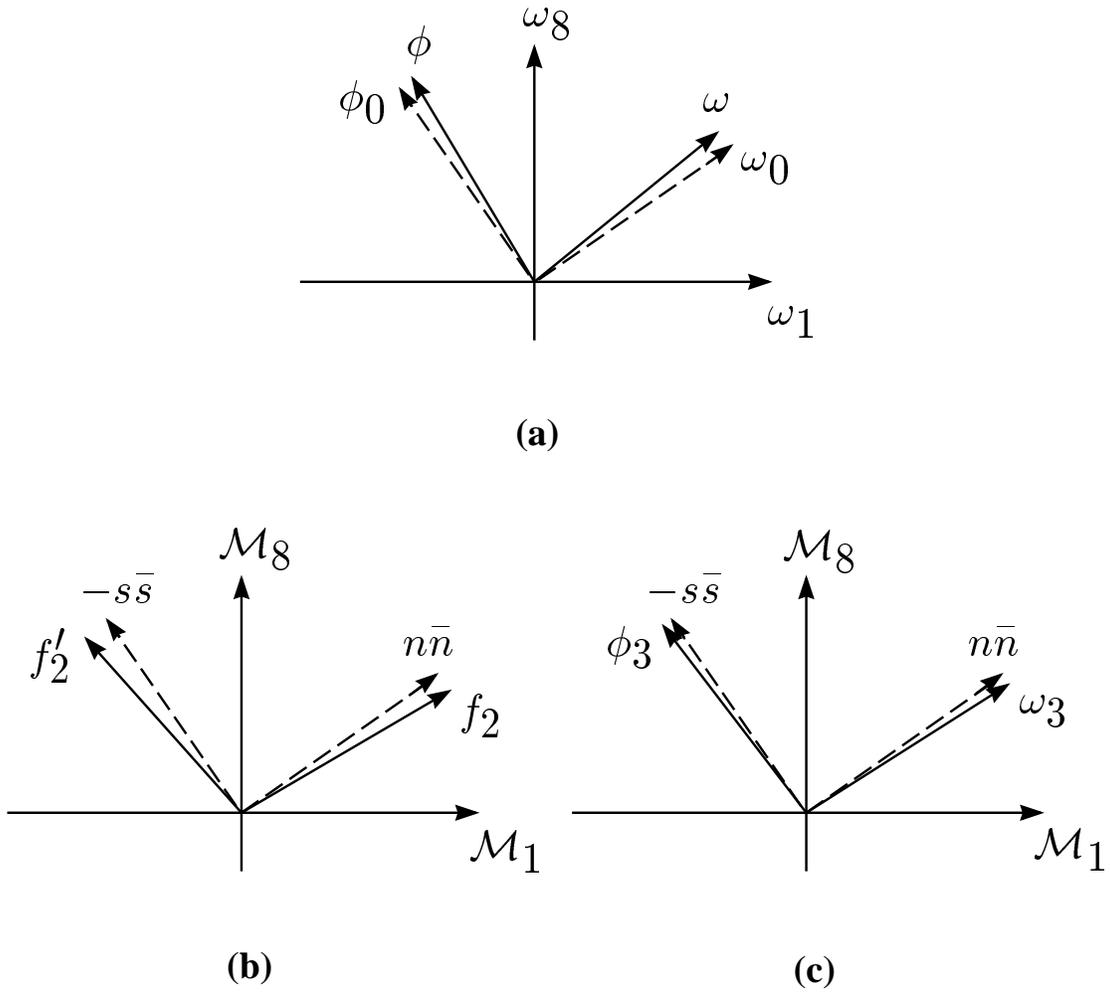}}}\fi
%
\vskip -0.2cm
  \caption{Physical isosinglet states vs. octet-singlet or ideally mixed states
  for (a) $1^{--}$, (b) $2^{++}$ and (c) $3^{--}$ mesons.
  The physical $\phi$ state would be on the wrong side of $\phi_0 = -s\bar s$ if one assumes
  equal rotation angles $\sin\theta_\phi = \sin\theta_\omega$.}
  \label{f2}
\end{figure}


\begin{thebibliography}{}
\bibitem{GI2} The third paper of Ref. \cite{GI} also discusses briefly the fact that
$\sin\theta_\omega \neq \sin\theta_\phi$ from the rescattering point of view.
\bibitem{peculiar} To be consistent with $\tan \delta_\omega =
 -T(\phi_0 \rightarrow \omega_0)/(m^2_{\phi_0}-m^2_{\omega_0})
> 0$, we find the somewhat peculiar result that $\phi(m_\omega) 
\sim \cos\delta_\omega\; s\bar s + \sin\delta_\omega\; n\bar n$ 
at the $\omega$ scale, i.e. mostly $s\bar s$ rather than $-s\bar s$.
In comparison, $\omega(m_\phi)
\sim \cos\delta_\phi\; n\bar n - \sin\delta_\phi\; s\bar s$.
These results can perhaps be tested by intereference experiments.
\end{thebibliography}
\end{document}